\documentclass[10pt,journal]{IEEEtran}

\usepackage{graphicx}
\usepackage{cite}
\usepackage{latexsym}
\usepackage{epsfig}
\usepackage{verbatim}
\usepackage{amssymb}
\usepackage[cmex10]{amsmath}
\usepackage{color}
\usepackage{url}
\usepackage{amsthm}
\usepackage{epstopdf}
\usepackage{epsfig}
\usepackage{amsthm}
\usepackage{epstopdf}
\usepackage{pseudocode}

\begin{document}

\newsavebox{\ieeealgbox}
\newenvironment{boxedalgorithmic}
  {\begin{lrbox}{\ieeealgbox}
   \begin{minipage}{\dimexpr\columnwidth-2\fboxsep-2\fboxrule}
   }
  {
   \end{minipage}
   \end{lrbox}\noindent\fbox{\usebox{\ieeealgbox}}}

\title{Joint User Association and Resource Allocation in Heterogeneous Cellular Networks: \textcolor{black}{Comparison of Two Modeling Approaches}}
\author{Dariush~Fooladivanda, and Catherine Rosenberg, ~\IEEEmembership{Fellow, ~IEEE}

\thanks{The first author is with the Department of Mechanical and Aerospace Engineering, University of California San Diego, and the second author is with the Department of Electrical and Computer Engineering, University of Waterloo. (Email: dfooladi@ucsd.edu, cath@uwaterloo.ca).}}

\newtheorem{thm}{Theorem}
\newtheorem{prop}{Proposition}
\newtheorem{lem}{Lemma}
\newtheorem{cor}{Corollary}
\newtheorem{defn}{Definition}
\newtheorem{assumption}{Assumption}
\newtheorem{rem}{Remark}
\thispagestyle{empty}

\maketitle
\begin{abstract}
The performance of different combinations of user association (UA) and resource allocation (RA) in heterogeneous cellular networks has been extensively studied using a classic modeling approach based on system snapshots. There have been also many studies focusing on the dynamics of the system \textcolor{black}{using queueing models}. These modeling approaches are rarely compared with each other though they each bring different insights to the design problem. In this paper, we consider a queueing model-based approach to study the interplay of UA and RA, and compare the results to those obtained using snapshot models. Specifically, we formulate three different joint UA and
RA optimization problems corresponding to the following three performance metrics: the maximum achievable arrival rate, the average system delay, and the maximum per-user delay. These problems are non-convex integer programs. We have therefore developed numerical techniques to compute either their exact solutions or tight lower bounds. We obtain results for different combinations of RA and UA schemes, and compare the trends with those obtained via the snapshot approach. The trends on RA are very similar, which we take as a cross-validation of the two modeling approaches for this kind of problem. The trends on user association are somewhat different which indicates a lack of robustness of the results and the need for a careful validation of UA models.
\end{abstract}

\IEEEpeerreviewmaketitle

\section{Introduction}
This paper focuses on the downlink of heterogeneous cellular networks (Hetnets) composed of macro base stations (BS) overlaid with a wide range of low-power BSs such as picos, femtos, and relays, creating small cells that are designed to improve coverage as well as spectral efficiency per unit area~\cite{low_power}, \cite{3GPP}, \cite{specteff}. We consider \textcolor{black}{orthogonal frequency-division multiplexing (OFDM) based Hetnets} (e.g., LTE-A) and hence sub-channels are the resources to allocate among the available BSs in the system. A resource allocation scheme determines how to allocate the sub-channels among the BSs, the user association policy defines a set of
rules for assigning users to the different BSs in the system, and the scheduling at each BS determines how to use the power budget on the allocated sub-channels at each BS and how to share the resources among the associated users. The choice of a RA, a UA, and a scheduling scheme determines the amount of interference \textcolor{black}{seen by each user} as well as the downlink received signal strength for each (BS, associated user) pair. Note that a user's throughput is not only a function of the number of sub-channels available at the BS and the level of interference, but is also a function of \textcolor{black}{the modulation and coding schemes as well as the other users associated with the same BS}. A decision to associate a user with one BS will affect the throughput seen by that user, as well as the throughput seen by the other users associated with that BS.

The performance of different combinations of RA, UA, and scheduling schemes in Hetnets has been extensively studied using a classic modeling approach in \cite{cell_asso}, \cite{df}, \cite{wei1}. In this modeling approach, a snapshot of the system is studied assuming that there are $N$ \emph{greedy users} placed at random in the system area, and that each BS has an infinite backlog of packets for each of its users. This modeling approach enables the formulation of many very detailed network utility maximization problems, and the evaluation of the throughput performance of various combinations of scheduling, fairness criteria, power control, UA, and RA schemes over a large number of \emph{independent} snapshots of the system. We call this approach the \emph{snapshot approach}.

In practical cellular systems, users enter the system, download a file (or visit a few web-pages), and leave the system when the file has been downloaded (or when the web-pages have been visited). Such users want to download their files as fast as possible. Such a practical system can certainly be represented by a sequence of snapshots. However, these snapshots are \emph{correlated} to each other by the dynamics of the system. The sequence of snapshots is also highly dependent on the deployed UA, RA, and scheduling. For example, a badly engineered system will keep the users longer in the system, and hence a new arrival will see a typical state that has much more users in the system than a well-engineered one. Therefore, we wonder if the conclusions drawn out of the snapshot approach are robust, i.e., if we would get similar results by evaluating different combinations of scheduling, power control, UA, and RA schemes in a more dynamic setting.

\textcolor{black}{While there has been several studies stressing the dynamic aspects of the problem, we are not aware of a study that compares the modeling approaches on a fair ground. This is one of the two objectives of the paper. To do so, we had to propose a queueing-based optimization framework for the joint UA, RA and scheduling. We call this approach the \emph{queueing-based approach}. This framework captures the dynamics in users' arrival and service times, and takes into account the user association and the resource allocation assuming the scheduling is proportional fair. More precisely, \textcolor{black}{we model the coverage area of each BS as a multi-class processor-sharing (PS) queue, and hence the Hetnet can be seen as a set of PS queues.} Our other objective is to study, using this framework, the long-run performance of the Hetnet, for different combinations of RA schemes and UA policies.}

Note that the \emph{snapshot approach} and the \emph{queueing-based approach} model the system under different sets of assumptions.
\textcolor{black}{Hence, our objective is to \emph{qualitatively} compare  the trends obtained via the two approaches.}



Our contributions are as follows:
\begin{enumerate}
\item  We formulate an offline tractable \textcolor{black}{queueing-based optimization} framework to analyze and compare different combinations of UA and RA schemes for the downlink of Hetnets. This framework allows us to study scenarios in which the arrival rate of users into the system is spatially homogeneous or in-homogeneous to model hot-spots. We consider three RA schemes (\emph{Co-channel deployment} (CCD), \emph{Orthogonal deployment} (OD), and \emph{Partially shared deployment} (PSD)). Given a resource allocation scheme, we formulate joint user association and resource allocation optimization problems corresponding to the optimization of the following three
criteria: The first one corresponds to the maximum achievable \textcolor{black}{user arrival rate} that the system can handle (i.e., the maximum arrival rate for which all the queues\footnote{\textcolor{black}{There is one queue per base station.}} are stable). The second criterion is the average system delay while the third one is the maximum per-user delay.
\item These problems are non-convex integer programs. We develop numerical techniques to compute the exact solutions for two of them and tight lower bounds for the other one.
\item We perform a thorough performance analysis of different combinations of RA schemes and UA policies \textcolor{black}{using this queueing framework.}
\item We provide a thorough comparison of the engineering insights obtained via this modeling approach to those obtained from the \emph{snapshot approach}, and show that the engineering insights on RA schemes are consistent while those on UA rules are not.
\end{enumerate}

The paper is organized as follows: Section \ref{sec_lit} presents an overview of the related works. The system model is introduced in Section \ref{sys_dynamic}. In Section \ref{pb_formulation}, \textcolor{black}{we formulate three joint user association and resource allocation optimization problems for \emph{OD} that differ in terms of their objective functions} \textcolor{black}{and since there are mixed integer non-convex programs, we propose  different solution techniques to solve them in Section~\ref{solution_tech}}. We provide numerical results along with some engineering insights in Section \ref{numerical_dynamic}, and compare the \emph{snapshot} and \emph{queueing-based} approaches in Section \ref{comp_App}. Section~\ref{concl} concludes the paper. All the proofs are presented in the Appendix.

\section{Literature Background}\label{sec_lit}
The performance of different resource allocation and user association schemes in Hetnets has been extensively studied \cite{dv1}, \cite{ref8} using the \emph{snapshot} approach. A comprehensive overview of the proposed user association and resource allocation schemes is provided in \cite{a_survey_3GPP}.

In \cite{df}, Fooladivanda \textit{et al.} study the interplay of user association and resource allocation for the downlink of a Hetnet that consists of a macro
BS and many pico BSs using a snapshot of the system and assuming that
greedy users. They select proportional fairness (PF) as their global objective function, and formulate joint optimization problems that are non-convex integer programs. Then, they develop techniques to obtain
upper bounds on the system's performance. They use these upper bounds to quantify how
well different combinations of UA rules (Small-cell First \cite{pimrc2011}, Range Extension \cite{qual}, and \textcolor{black}{signal-to-noise-plus-interference ratio (SINR) based}) and
RA schemes (CCD, OD, and PSD \cite{a_survey_3GPP}) perform in
Hetnets. While several studies have used the \emph{snapshot approach} to analyze different combinations of UA and RA schemes in Hetnets, there are also several studies that analyze Hetnets in a dynamic setting \cite{ref_stability_1}, \cite{ref_DA_1}.

In \cite{dv2}, Kim \textit{et al.} consider the system area of a general multi-cell wireless system as a continuous space where users arrive randomly, download files, and leave after being served. The authors capture the dynamics in users' arrival and service times with a multi-class PS queueing model. They consider infinitely many classes since each user can arrive at any point in the system area, and formulate a UA problem with a generic $\alpha$-fair objective function on the load of each cell. This problem has an infinite number of constraints, and is computationally intractable. The authors propose and analyze an iterative distributed user association policy that converges to a global optimum under a set of assumptions. Finally, they propose admission control policies for the scenario where the system is overloaded and cannot be stabilized. In \cite{ref_14_3} and \cite{ref_14_4}, the authors use the framework developed in \cite{dv2}, and propose energy-efficient user association rules.

\textcolor{black}{In \cite{dyn_New1}, the authors model the dynamics of a Hetnet using a queueing model, and propose two spectrum allocation schemes. They propose efficient algorithms for computing optimal spectrum allocations, and show numerically that the proposed schemes significantly outperform orthogonal and full-frequency reuse
allocations under all traffic conditions. In \cite{dyn_New2}, the authors propose a tractable approach to analyze delay in Hetnets with spatio-temporal random arrival of traffic, and evaluate the effect of different scheduling policies on the delay performance. The authors numerically show that the delay performance of round-robin scheduling outperforms first-in first-out scheduling for heavy traffic while the reverse is true for light traffic.}

\textcolor{black}{Using stochastic geometry, the authors in \cite{dyn_New3} develop an analytical framework for an accurate
prediction of the flow-level performance of multi-tier networks. They derive analytically the per flow delay, load, and congestion probability of BSs for different tiers. They apply their model to a 2-tier network based on LTE and WiFi, and study the performance of
different user association rules.}

\textcolor{black}{In this study, we focus on joint resource allocation and user association as well as on how two different modeling approaches compare RA policies and UA schemes. Extensive work has been done on UA and RA using a dynamic approach (not always using a queueing model). However, none of these works can be used to compute the optimal UA and RA, and to obtain engineering insights on the performance of different combinations of UA and RA schemes. We are also not aware of works that compare different modeling techniques to address the problems of UA and RA in Hetnets.}


\section{System Model}\label{sys_dynamic}
\begin{figure}[t]
\begin{center}
\includegraphics[width=3.5in]{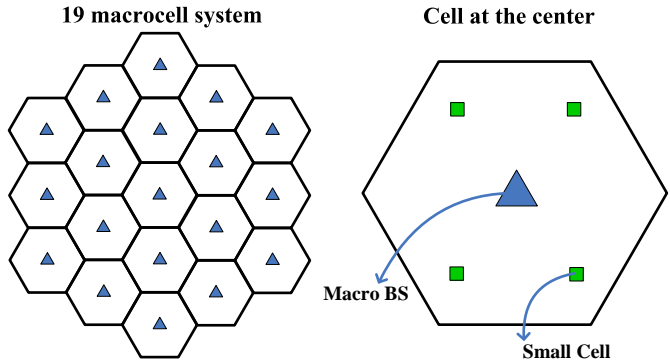} \caption{A Hetnet comprising 19 macro BSs (the triangles). Each macro cell $j$ is overlaid with $B_j$ small cells (the squares). Small cell locations for the cell at the center are shown in the right-hand side figure.} \label{system_model162}
\end{center}
\end{figure}

\textcolor{black}{We consider a multi-tier communication system composed of $m$ macro cells, (see Fig. \ref{system_model162}). Each macro cell $j$ is overlaid with $B_j$ small cells that are identical in terms of transmit power, antenna gain, and backhaul capacity. Let $\mathcal{M}$ and $\mathcal{SC}$ denote the sets of macro and small cell BSs in the system, respectively.} The system is an OFDM system with $rM$ sub-channels, each of bandwidth $b$. These sub-channels are divided among the macro BSs based on conventional frequency reuse \cite{Goldsmith}, i.e., given reuse factor $r$, each macro BS is granted one of the $r$ groups of $M$ sub-channels.

\textcolor{black}{We discretize the set of locations at which users can be in order to obtain tractable optimization problems. Note that modeling the system area of a wireless system as a continuous space will result in formulations that are computationally intractable in a centralized fashion\cite{dv2}.} Let $\mathcal{L} = \{1,\ldots,L\}$ denote the set of possible user locations within the system area. We focus on the downlink, and make the following assumptions:
\begin{itemize}
\item[A.1] The maximum transmit powers of the macro ($\widehat{P}_m$) and small cells ($\widehat{P}_p$) are fixed and known a priori.
\item[A.2] Each small cell is connected to the macro BS via a high capacity wired backhaul.
\item[A.3] Each user associates with a single BS\footnote{\textcolor{black}{The generic term BS refers to both a macro cell or a small cell.}}.
\item[A.4] Users arrive at location $i\in\mathcal{L}$ according to a Poisson process with density ${\lambda}_i=\alpha_i \lambda$ where $0 \leq \alpha_i \leq 1$ and $\sum_{i=1}^L{\alpha_i}=1$, i.e., the arrival rate into the system is given by the vector $\overline{\lambda}= \lambda(\alpha_1,\cdots,\alpha_L)$ where the vector $\overline{\alpha}=(\alpha_1,\cdots,\alpha_L)$ represents \textcolor{black}{the spatial in-homogeneity of the} \textcolor{black}{traffic distribution over the system area. The case $\alpha_j=1/L$ represents the homogeneous case. We assume that $\overline{\alpha}$ is given, and by a slight abuse of notation, we will call $\lambda$ the user arrival rate.}
\item[A.5] Users arriving to the system download files whose sizes are independent and identically distributed (i.i.d.) random variables of mean $F$ bits.
\item[A.6] Users depart the system as soon as their files have been downloaded completely.
\end{itemize}

\subsection{Resource Allocation}
Sub-channels\footnote{We use the terms channel and sub-channels interchangeably.} are \textcolor{black}{the resources that we allocate to the different BSs. We have already assumed that each macro cell receives $M$ sub-channels. We assume that they all use the same scheme to share their channels between the MBS and the SCs.} We consider three different resource allocation schemes:
\begin{itemize}
\item  \emph{Co-channel deployment (\emph{CCD}):} The macro and small cells transmit on all the $M$ sub-channels.
\item \emph{Orthogonal deployment  (\emph{OD}):} $K$ sub-channels are dedicated exclusively to the pool of
small cells and the remaining $(M-K)$ sub-channels are dedicated to the macro BS. Each small cell transmits on all the $K$ sub-channels.
\item \emph{Partially shared deployment (\emph{PSD}):} $K$ sub-channels are shared by the macro and small cells, and the other $(M-K)$ sub-channels are dedicated to the macro BS. Each small cell transmits on all the $K$ sub-channels. The macro BS transmits on the $K$ channels with power budget $\widehat{P}_p$ and on the $(M-K)$ sub-channels with power budget $(\widehat{P}_m-\widehat{P}_p)$.
\end{itemize}

\subsection{Scheduling}
Power and time are the resources that the scheduler at each BS allocates to its users. \textcolor{black}{For \emph{CCD} and \emph{OD}, the scheduler allocates the power budget of the BS equally among all its allocated sub-channels while for \emph{PSD}, the scheduler at the macro BS allocates the power budget for each subset of sub-channels equally among all sub-channels
in the subset.} We \textcolor{black}{make the widely used and reasonable assumptions that:
\begin{itemize}
\item[A.7] Each BS uses a local proportional fair (PF) scheduling in which it transmits all the time on all its allocated sub-channels using equal power and allocates the same proportion of time to its users \cite{df}.
\item[A.8] The sub-channel gains are flat for each (BS, user) pair, and users do not move during their sessions. Hence, the channel gains do not change drastically during the lifetime of a session and are known.
\end{itemize}
Under these assumptions, the inter-cell interference remains the same irrespective of the load of each BS.} \textcolor{black}{Recall that we are only considering the downlink traffic in this paper.} There is then no load-coupling in our system. This assumption can be restrictive when the user arrival rate $\lambda$ is small since it is then possible that some BSs are not occupied, and hence, results for small $\lambda$ should not be over-interpreted (please see remark at the end of Section \ref{flow_d}. If the user arrival rate $\lambda$ is relatively large, all the BSs will have some users to serve. In addition, since our users are greedy, a BS will have to transmit all the time as soon as it has one user.

\subsection{Physical Link Model}
 Given a resource allocation scheme, we know the set of co-channel BSs and the transmit power of each BS on each of its sub-channels. We can then compute the SINR at location $i \in \mathcal{L}$ from BS $j\in \mathcal{M}\cup\mathcal{SC}$ on each sub-channel (call it $\gamma_{ij}$) as follows:
\begin{equation}\label{SINR_model_dynamic}\small
\gamma_{ij} = \frac{P_j~G_{ij}}{N_0+\sum_{h \in
\mathcal{I}_j}{P_h~G_{ih}}}
\end{equation}
where $\mathcal{I}_j$ is the set of BSs (macros and small cells) transmitting on the same set of sub-channels
(not including $j$) in the multi-tier system, $P_j$ is the transmit power of BS $j$
on a sub-channel, $N_0$ is the additive white Gaussian noise
power on the sub-channel, and $G_{ij}$ is the gain between
location $i$ and BS $j$ that accounts for the path loss, shadow fading,
antenna gain, and equipment losses.
Note that given a reuse factor $r$, and a resource allocation and its parameter (i.e., $K$ for OD and PSD), $\mathcal{I}_j$ as well as $P_j$ and $P_h$ can be determined.

Then, given the discrete \textcolor{black}{modulation and coding scheme (MCS) function $f(\cdot)$ that maps the SINR into a rate}, we can compute the link rate at location $i$ from BS $j$ on a sub-channel as $r_{ij} =f(\gamma_{ij})$ \cite{jg}.

\subsection{Flow-level Queueing Model}\label{flow_d}
We capture the system dynamics by a queueing model which takes into account the users' arrival and departure processes as well as the scheduling policy. \textcolor{black}{The system has $\sum_{j=1}^m (1+B_j)$ queues (one per BS) where $B_j$ is the number of SCs in macro area $j$\footnote{\textcolor{black}{In the case of PSD, macro $j$ has $B_j+2$ queues.}}}. \textcolor{black}{Recall that users arrive at location $i$ according to a Poisson process with rate ${\lambda}_i=\alpha_i \lambda$, and download files whose sizes are i.i.d. random variables of mean $F$.} Location $i$ might be in the coverage area of multiple BSs, and hence the users arriving to location $i$ have to decide to which BS to associate.
\textcolor{black}{\begin{itemize}
\item[A.9] We consider the class of UA schemes that make a decision based on physical layer parameters, i.e., based on our previous assumptions, users arriving at a certain location always associate with the same BS. Examples of UA schemes in this class are the one that selects the BS providing the highest SINR or small cell first \cite{pimrc2011}-\cite{RE}. This framework does not allow a UA rule that takes a decision based on the BS loads. Note that most practical UAs schemes are not load-based. 
\end{itemize}
Under this assumption, an UA policy defines a set of rules for assigning each location to a pre-defined set of BSs in the Hetnet. Hence, the UA determines the arrival rates at each queue. We can easily extent the framework to allow probabilistic UA, i.e., a user arriving at location $i$ that can hear say BSs $j_1$ and $j_2$,  joins $j_1$ with probability $p_i$ and $j_2$ with probability $1-p_i$ where $p_i$ is computed beforehand. For simplifying the notations and the derivations, we assume that each location is mapped to a unique BS in the following.}

Each BS \textcolor{black}{performs a local PF scheduling}, i.e., offers the same amount of time to all its users. Therefore, the users in each BS get served based on the processor sharing (PS) discipline \cite{Kleinrock1}. More precisely, we consider the coverage area of each BS as a generalized processor sharing queue, and each of the locations in the cell area as a class. Under our assumptions, each location $i$ (i.e., class $i$) sees a fixed SINR. Hence, each location has its own general service time distribution. \textcolor{black}{Since we assume that each BS transmits all the time and there is no coordination among the BSs, we can view the system, given a UA,  as a set of independent multi-class M/G/1 PS queues.}

\begin{rem}
\textcolor{black}{If we do not assume that the BSs transmit all the time, then the SINR's cannot be computed beforehand. Clearly, what we obtain with this assumption,  when $\lambda$ is small, is a lower bound on the performance since  better SINR's could be obtained when some cells are not transmitting. More precisely, our model provides a worst case scenario when $\lambda$ is small, i.e., an upper bound on the delay. Note that the low $\lambda$ regime is not a regime of great interest since the system is not under stress under this regime.
}
\end{rem}


\subsection{Performance Metrics}
In each of the multi-class M/G/1 PS queues, the per user service rate is a function of the number of channels available at the BS \textcolor{black}{(a function of the RA)}, the level of interference \textcolor{black}{(a function of the RA)}, and the current number of users associated with the BS \textcolor{black}{(a function of the UA)}. Hence, the deployed
UA and RA schemes will have a critical impact on service
rates and hence, on users' performance.

We focus on the long-run performance of the set of multi-class M/G/1 PS queues, and consider the following three performance metrics:
\begin{itemize}
\item \textit{Maximum Achievable Arrival Rate:}  Given a RA scheme $X$, a UA scheme $Y$, and an in-homogeneity vector $\overline{\alpha}$, let ${\lambda}_{max}(X,Y,\overline{\alpha})$ be the maximum user arrival rate for which the system is stable (i.e., all the queues are stable).

\item \textit{Average System Delay:} Given a RA scheme $X$ and a UA scheme $Y$, the average delay experienced by the users arriving to location $i$ (we call it the average delay of class $i$) depends on the service rate in location $i$ as well as the arrival rate vector \textcolor{black}{$\overline{\lambda}=\lambda\overline{\alpha}$}. The average delay over all classes in the system is what we call the average system delay.

\item \textit{Maximum Average Delay per Class:} Given a RA scheme $X$ and a UA scheme $Y$, the maximum average delay per class can be seen as a performance metric for edge users (i.e., users with low link rates).
\end{itemize}
\textcolor{black}{For more information on the performance metrics above, we refer the reader to \cite{queue_ref1}.}

Next, {\bf we focus on \emph{orthogonal deployment} (i.e., $X=\text{OD}$)}, and formulate three different joint user association and resource allocation optimization problems, one for each of the performance metrics. For \emph{partially shared deployment} and \emph{co-channel deployment}, we obtain similar problems that we do not present due to space limitations, \textcolor{black}{but we will present results on these RA schemes in Section \ref{numerical_dynamic}.}

\section{Problem Formulations for $X=\text{OD}$}\label{pb_formulation}
We first define the association variable $x_{ij}$ where $i\in\mathcal{L}$ and \textcolor{black}{$j\in \mathcal{M}\cup\mathcal{SC}$.} Let $x_{ij}=1$ if location $i$ is associated with BS $j$, and let it be $0$, otherwise. Hence, for all $i \in \mathcal{L}$, $\sum_{j \in \mathcal{M}\cup\mathcal{SC}} x_{ij}=1$.


\subsection{Maximizing the Achievable Arrival Rate}\label{stability_sec}
Given \textcolor{black}{the system and }a vector $\overline{\alpha}$, \textcolor{black}{our objective is to maximize the feasibility region, i.e., the achievable user arrival rate over all values of $K$, the RA parameter, and all UA's \textcolor{black}{in the class of UAs under study (see A.9)}. Let ${{\lambda}}_{max}^{\star}(\overline{\alpha})$ be this maximum.} As mentioned earlier, it is linked to the stability of the multi-class M/G/1 PS queues serving the system area. Given the $(x_{ij})$'s, the queues are independent, and then, the system is stable \emph{if and only if} each one of the queues is stable. Define the load factor of BS $j$ (call it $\rho_j$) as follows:
\[\rho_j = \sum_{i\in\mathcal{L}}{x_{ij}  \frac{{\alpha}_i {\lambda} F}{K_j r_{ij}}}\]
where $K_j$ denotes the number of sub-channels allocated to BS $j$, i.e., \textcolor{black}{$K_j=M-K$ if $j\in\mathcal{M}$, and $K_j=K$ otherwise}. Recall that the per-channel link rates $r_{ij}$'s can easily be computed under our assumptions given the $K_j$'s.

The system is stable \emph{if and only if} the load factor of each queue is strictly less than one \cite{queue_ref1}, i.e.,
\begin{equation}\label{stability_cond}{\lambda} \sum_{i\in\mathcal{L}}{x_{ij}  \frac{\alpha_i F}{K_j r_{ij}}} <1~,\quad \forall j \in \mathcal{M}\cup\mathcal{SC}.\end{equation}

Given the RA parameter $K$ and the vector $\overline{\alpha}$, an arrival rate ${\lambda}$ is said to be feasible \emph{if and only if} there exists an user association $\{x_{ij}\}$ for which (\ref{stability_cond}) is satisfied. By using (\ref{stability_cond}), we can easily check the feasibility of ${\lambda}$ for a given UA, but it is harder to find whether there exists an user association for which (\ref{stability_cond}) is satisfied. In order to write tractable optimization problems, we prefer not to work with strict inequalities, and hence we introduce a parameter $\bar{\rho}$, and assume that the load at each BS $j$ cannot be larger than $\bar{\rho}$ (i.e., $\rho_j \le \bar{\rho}$) where $0<\bar{\rho}<1$ is a constant and can be made arbitrarily close to one. Therefore, for a given $0<\bar{\rho}<1$, we replace the stability condition (\ref{stability_cond}) by
\begin{equation}\label{stability_cond1}{\lambda} \sum_{i\in\mathcal{L}}{x_{ij} \frac{\alpha_i F}{K_j r_{ij}}} \le \bar{\rho}~,\quad \forall j \in \mathcal{M}\cup\mathcal{SC}.\end{equation}

Our objective is to maximize the  achievable arrival rate for a given vector $\overline{\alpha}$ over all $K$ and all UA's. To compute this metric,
we formulate a joint user association and resource allocation problem in which the variables are $K$, $\{x_{ij}\}$, and ${\lambda}$. The problem can be formulated as follows: Given the OD resource allocation, the channel gains, the rate function $f(\cdot)$, the vector $\overline{\alpha}$, the average file size $F$, and $\bar{\rho}$, compute $K$, $\{x_{ij}\}$, and ${\lambda}$ so
as to maximize the maximum achievable arrival rate:
\begin{subequations}
\begin{align}
{\textbf{P}}_{s}&:~\max_{\{x_{ij}\},{\lambda},K}  \quad {{\lambda}}\nonumber\\
\label{stability_cond_pb} & {\lambda} \sum_{i\in\mathcal{L}}{x_{ij} \frac{\alpha_i F}{K_j r_{ij}}} \le \bar{\rho}~,~~~ \forall j \in \mathcal{M}\cup\mathcal{SC}\\
\label{lolo} & \gamma_{ij} = \frac{P_j~G_{ij}}{N_0+\sum_{h \in \mathcal{I}_j}{P_h~G_{ih}}} ~\forall i \in \mathcal{L}, \forall j \in \mathcal{M}\cup\mathcal{SC}\\
\label{pp1} & \textcolor{black}{P_j = \frac{\widehat{P}_p}{K_j}, K_j=K,~ \forall j \in \mathcal{SC}}\\
\label{pp2} & \textcolor{black}{P_j = \frac{\widehat{P}_m}{K_j}, K_j=M-K,~ \forall j \in \mathcal{M}}\\
\label{channel3} &  r_{ij} =f(\gamma_{ij}) ~\forall i \in \mathcal{L}, \forall j \in \mathcal{M}\cup\mathcal{SC}\\
\label{UA_const_pb}&\sum_{j \in \mathcal{M}\cup\mathcal{SC}} x_{ij}=1~, \quad \forall i  \in \mathcal{L}\\
\label{channel1} &K\in\{1,2,\cdots,M\}, \\
\label{integer}{\lambda} &\ge0, x_{ij} \in \{0,1\}, ~\forall i \in \mathcal{L}, \forall j \in \mathcal{M}\cup\mathcal{SC}
\end{align}
\end{subequations}

The proposed joint user association and resource allocation optimization problem enables us to compute the maximum achievable arrival rate ${{\lambda}}_{max}^\star(\overline{\alpha})$ where the star refers to the fact that this is a maximum over all possible UA's. Arrival rate ${\lambda}$ can be as large as ${{\lambda}}_{max}^\star(\text{OD},\overline{\alpha})$ if we use the user association $\{x_{ij}^\star\}$ where $\{x_{ij}^\star\}$ is the solution to ${\textbf{P}}_{s}$; otherwise, the system is not necessarily stable. For any other UA scheme $Y$ defined by $\{x_{ij}^{y}\}$, the maximum achievable arrival rate (call it ${{\lambda}}_{max}(Y,\overline{\alpha}))$ can be easily computed by using (\ref{stability_cond1}), and is less than or equal to ${{\lambda}}_{max}^\star(\overline{\alpha})$. \textcolor{black}{Note that ${\textbf{P}}_{s}$ is a mixed integer non-convex program that is NP-hard and cannot be solved  efficiently as is. We will propose a solution technique to solve it in Section~\ref{solution_tech}.}


Next, we focus on the average system delay and the maximum average delay per class, and formulate  joint user association and resource allocation optimization problems assuming $\overline{\alpha}$, ${{\lambda}}_{max}^\star(\overline{\alpha})$, and an arrival rate ${\lambda} \le {{\lambda}}_{max}^\star(\overline{\alpha})$ are given (i.e., the system is in its stability region).

\subsection{Delay-based Metrics}
\textcolor{black}{Let the average delay per class $i$ (i.e., per location)  be $T_i$, and the average system delay be $T$. Then, we have:}

\begin{equation}\label{delai2} T =\left({\sum_{i\in \mathcal{L}}{{\lambda}_i T_i}}\right)/\left({\sum_{i\in \mathcal{L}}{{\lambda}_i} }\right)\end{equation}

\textcolor{black}{Note that, given $K$ and a user association scheme $Y$ defined by $\{x_{ij}^{y}\}$, we can compute the average delay at location $i$  by \cite{queue_ref1}}
\begin{equation}\label{delay_per_loc_def}T_i^y= \sum_{j\in \mathcal{M}\cup\mathcal{SC}}{x_{ij}^{y} \frac{F}{(1-\rho_j^{y}) K_j r_{ij}}}\end{equation}
\textcolor{black}{where $\rho_j^y$ is the load factor of BS $j$ given UA scheme $Y$.}

\textcolor{black}{Given $\overline{\alpha}$ and ${\lambda} \le {{\lambda}}_{max}^\star(\overline{\alpha})$, we will formulate two problems, one whose objective is to  min-max the average delay  per class $T_i$ over all $i$'s (i.e., all locations) and the second is to minimize the average system delay $T$. The variables for these problems are $K$ and the UA variables $\{x_{ij}\}$.}

\textcolor{black}{For space reason, we present the two problems at once, i.e., we formulate a generic problem} ${\textbf{P}}_{delay}(q)$ where problem ${\textbf{P}}_{delay}(1)$ is the min-max problem and ${\textbf{P}}_{delay}(2)$ is the other problem. Let $d_1(\{T_i\})=\max_{i \in \mathcal{L}}{T_i}$ and $d_2(\{T_i\})=T$, then, the generic problem ${\textbf{P}}_{delay}(q)$ can be formulated as follows: given the channel gains, the rate function $f(\cdot)$, the vector $\overline{\alpha}$, the average file size $F$, $\bar{\rho}$, and ${\lambda}$:
\begin{subequations}
\begin{align}
{\textbf{P}}_{delay}(q):~~&\min_{\{x_{ij}\},\{\rho_j\},\{T_i\},K }  \quad {d_q(\{T_i\})}\nonumber\\
\text{subject to}~&(\ref{stability_cond_pb})-(\ref{channel1})\nonumber\\
\label{delay_per_loc_pb}T_i &= \sum_{j\in \mathcal{M}\cup\mathcal{SC}} {x_{ij} \frac{F}{(1-\rho_j) K_j r_{ij}}}, \forall i \in\mathcal{L} \\
&\label{rho_pb}\rho_j={\lambda} \sum_{i\in\mathcal{L}}{x_{ij} \frac{\alpha_i F}{K_j r_{ij}}},~ \forall j \in \mathcal{M}\cup\mathcal{SC}\\
&\label{x_pb}x_{ij} \in \{0,1\}~, \quad \forall i \in \mathcal{L},~ \forall j \in \mathcal{M}\cup\mathcal{SC}
\end{align}
\end{subequations}
\textcolor{black}{Note that ${\textbf{P}}_{delay}(q)$  is a mixed integer non-convex program that is NP-hard and cannot be solved  efficiently as is. We propose  two different solution techniques to solve the problem for $q=1$ and $q=2$  in Section~\ref{solution_tech}.}

\section{Solution Techniques}\label{solution_tech}
In the problems defined above,  some variables such as $K$ and $\{x_{ij}\}$ are discrete while some others such as $\{\rho_j\}$ and $\{T_i\}$ are continuous. In addition, the rate function $f(\cdot)$ is a discrete function (i.e., a non-differentiable function). Since the variable $K$ takes one of the integer values in $\{1,2,\cdots,M\}$, exact solutions to ${\textbf{P}}_{s}$ and ${\textbf{P}}_{delay}(q)$ can be obtained by solving these problems iteratively for all possible values of $K$, and then selecting the best solution.

Let ${\textbf{P}}_{s}(K)$ and ${\textbf{P}}_{delay}(q,K)$ for $q=1,2$ be the problems obtained by fixing the resource allocation parameter $K$. These problems are still non-convex integer programs. Let ${{\widehat{\lambda}}}_{max}(K,\overline{\alpha})$ (resp. $D_{q}(K,\overline{\alpha})$) denote the optimal value of the objective function in ${\textbf{P}}_{s}(K)$ (resp. ${\textbf{P}}_{delay}(q,K)$). The exact solution to ${\textbf{P}}_{s}$ (resp. ${\textbf{P}}_{delay}(q)$) can be obtained by solving $\max_{K}{\{{{\widehat{\lambda}}}_{max}(K,\overline{\alpha})\}}$ (resp. $\min_{K}{\{D_{q}(K,\overline{\alpha})\}}$).

\subsection{Maximum Achievable Arrival Rate}
To obtain an exact solution to ${\textbf{P}}_{s}(K)$, we first formulate a new user association problem called ${\textbf{P}}_{s}'(K)$, and then show that an optimal solution to ${\textbf{P}}_{s}(K)$ can be obtained by solving ${\textbf{P}}_{s}'(K)$. We define ${\textbf{P}}_{s}'(K)$ as follows: Given $K$, $\overline{\alpha}$, and $F$:
\begin{subequations}
\begin{align}
{\textbf{P}}_{s}'(K):~~&\min_{\{x_{ij}\},\Lambda}  \quad {\Lambda}\nonumber\\
\text{subject to}~&\nonumber\\
& \sum_{i\in\mathcal{L}}{x_{ij} \frac{\alpha_i F}{K_j r_{ij}}} \le \Lambda~,~\forall j \in \mathcal{M}\cup\mathcal{SC}\nonumber\\
&\sum_{j \in \mathcal{M}\cup\mathcal{SC}} x_{ij}=1~, ~ \forall i  \in \mathcal{L}\nonumber\\
&\Lambda \ge0,~x_{ij} \in \{0,1\}~, ~ \forall i \in \mathcal{L},~ \forall j \in \mathcal{M}\cup\mathcal{SC}\nonumber
\end{align}
\end{subequations}
where \textcolor{black}{all $r_{ij}$'s and $K_j$'s are computed beforehand} (since $K$ is given) and used as input parameters to ${\textbf{P}}_{s}'(K)$.

The following result shows that an exact solution to ${\textbf{P}}_{s}(K)$ can be obtained by solving ${\textbf{P}}_{s}'(K)$. The proof is provided in the appendix.

\begin{thm}\label{lem_lambda}
Given $\bar{\rho}$, $\overline{\alpha}$, $F$, and $K$, the solution to ${\textbf{P}}_{s}(K)$, ${{\widehat{\lambda}}}_{max}(K,\overline{\alpha})$, is equal to $\left({\bar{\rho}}/{\Lambda^\star}\right)$ where $\Lambda^\star$ denotes the optimal solution to ${\textbf{P}}_{s}'(K)$.
\end{thm}

We can now work with ${\textbf{P}}_{s}'(K)$ which is an integer linear program and can be solved with a commercial solver. The proposed technique enables us to compute the maximum achievable arrival rate ${{\widehat{\lambda}}}_{max}(K,\overline{\alpha})$ for all possible values of $K$, and then selecting the largest one over all $K$'s to find ${{\lambda}}_{max}^\star(\text{OD},\overline{\alpha})$. Next, we focus on the maximum average delay per class.

\subsection{The Min-Max Problem On The Average Delay per Class}
We can obtain an exact solution to ${\textbf{P}}_{delay}(1,K)$ by using the following two steps. In the first step, given $t>0$, we formulate a feasibility problem called ${\textbf{P}}'_{delay}(t)$, and show that the optimal value of the objective function in ${\textbf{P}}_{delay}(1,K)$ is less than or equal to $t$ if the problem ${\textbf{P}}'_{delay}(t)$ is feasible. The feasibility problem ${\textbf{P}}'_{delay}(t)$ is an integer linear program which can be solved with a commercial software. In the second step, we propose an iterative algorithm which solves a limited number of instances of ${\textbf{P}}'_{delay}(t)$ to obtain an exact solution to ${\textbf{P}}_{delay}(1,K)$.

\noindent \textbf{STEP~1 :} We formulate a feasibility problem called ${\textbf{P}}'_{delay}(t)$ as follows: Given $t>0$, $\bar{\rho}$, ${\lambda}$, $\overline{\alpha}$, $K$, and $F$:
\begin{subequations}
\begin{align}
{\textbf{P}}^{\prime}_{delay}(t):~~&\min_{\{x_{ij}\},\{\rho_j\}}  \quad {\mathbf{1}}\nonumber\\
\text{subject to}~&(\ref{stability_cond_pb}),(\ref{UA_const_pb}),~(\ref{rho_pb})-(\ref{x_pb})\nonumber\\
t (1&-\rho_j) \ge {x_{ij} \frac{F}{K_j r_{ij}}}~\forall i \in\mathcal{L}, \forall j\in \mathcal{M}\cup\mathcal{SC}\nonumber
\end{align}
\end{subequations}
where all $r_{ij}$'s and $K_j$'s can be computed beforehand (since $K$ is given) and used as input parameters.

The following result shows that we can check whether the optimal value of ${\textbf{P}}_{delay}(1,K)$ is less than or equal to a given value $t$ by solving the feasibility problem ${\textbf{P}}^{\prime}_{delay}(t)$. A sketch of the proof is provided in the appendix.

\begin{thm}\label{delay_1_lem}
Given $\bar{\rho}$, ${\lambda}$, $\overline{\alpha}$, $K$, and $F$, let $p^\star$ denote the optimal value of the objective function in ${\textbf{P}}_{delay}(1,K)$. If the feasibility problem ${\textbf{P}}^{\prime}_{delay}(t)$ is feasible for a given value $t>0$, then we have $p^\star \le t$; otherwise, we have $p^\star>t$.
\end{thm}

Based on Theorem \ref{delay_1_lem}, we propose an algorithm to compute an optimal solution to ${\textbf{P}}_{delay}(1,K)$.

\noindent \textbf{STEP~2 :} To obtain an exact solution to ${\textbf{P}}_{delay}(1,K)$, we first compute a feasible solution to ${\textbf{P}}_{delay}(1,K)$. \textcolor{black}{To do so, we only need to compute a user association $\{x_{ij}\}$ that satisfies the constraint in (\ref{stability_cond_pb}) and (\ref{UA_const_pb}). Note that we can easily compute such feasible solutions since the constraint in (\ref{stability_cond_pb}) and (\ref{UA_const_pb}) are linear.} Let $t_0$ denote the value of $\max_{i \in \mathcal{L}}{T_i}$ for that association rule. Given $t_0>0$ and ${\lambda}$, we start with the interval $I_0=[0,t_0]$. Clearly, the interval $I_0$ contains the optimal value of the objective function in ${\textbf{P}}^\prime_{delay}(1,K)$ since $t_0$ is a feasible delay and an upper bound on the optimal delay. We solve the feasibility problem ${\textbf{P}}^{\prime}_{delay}(t)$ at the midpoint of $I_0$, i.e., $t=\frac{t_0}{2}$. This determines whether the optimal value $p^\star$ is in the lower or upper half of $I_0$. We then obtain a new interval which contains the optimal value $p^\star$. Note that the width of the new interval is reduced to half of the interval in the previous iteration. We repeat this process until the width of the interval is sufficiently small. In each step, the width of the interval is reduced by two folds, and hence after $k$ iterations, the length of the interval is $2^{-k}t_0$. Therefore, we need $\lceil\text{log}_2(\frac{t_0}{\epsilon})\rceil$ iterations to obtain the optimal value of ${\textbf{P}}_{delay}(1,K)$ with the desired precision $\epsilon$. A formal description of the proposed algorithm is given in Algorithm \ref{h_01}.

\begin{figure}
\begin{pseudocode}{The proposed algorithm }{t_0,\epsilon}
\label{h_01}
\mbox{\noindent\rule{7cm}{0.4pt}}\\
\mbox{1: {\bf Initialize:} $\ell=0$, $u=t_0$, $i=0$}\\
\mbox{2: {\bf Repeat}}\\
\mbox{3: $\quad$ Set $t_i=\frac{\ell+u}{2}$}\\
\mbox{4: $\quad$ Solve the feasibility problem ${\textbf{P}}^{\prime}_{delay}(t_i)$.}\\
\mbox{5: $\quad$ {\bf If} ${\textbf{P}}^{\prime}_{delay}(t_i)$ is feasible} \\
\mbox{$\quad$ $\quad$ $\quad$ Set $u=t_i$}\\
\mbox{~~  $\quad$ {\bf Else}}  \\
\mbox{$\quad$ $\quad$ $\quad$ Set $\ell=t_i$}\\
\mbox{6: $\quad$ Set $i=i+1$}\\
\mbox{7: {\bf Until} $u-\ell \le \epsilon$ }\\
\mbox{\noindent\rule{7cm}{0.4pt}}
\end{pseudocode}
\end{figure}

In summary, the problem ${\textbf{P}}_{delay}(1,K)$ is a mixed integer nonlinear problem which cannot be solved for relatively large networks (i.e., $L$ relatively large). The proposed algorithm enables us to solve ${\textbf{P}}_{delay}(1,K)$ with the desired precision $\epsilon$ by solving a limited number of linear integer programs which can be solved with a commercial solver.

\subsection{Minimizing The Average System Delay}
 Our goal is to obtain a tight lower bound on the optimal value of the objective function in ${\textbf{P}}_{delay}(2,K)$. To do so, we first formulate a new user association problem called ${\textbf{Q}}_{delay}$ as follows: Given $\bar{\rho}$, ${\lambda}$, $K$, $\overline{\alpha}$, and $F$:
\begin{subequations}
\begin{align}
{\textbf{Q}}_{delay}:~~&\min_{\{x_{ij}\},\{\rho_j\} }  \quad {\sum_{j \in \mathcal{M}\cup\mathcal{SC}}{\frac{1}{1-\rho_j}}}\nonumber\\
\text{subject to}~&(\ref{stability_cond_pb}), (\ref{UA_const_pb}), (\ref{rho_pb})\nonumber\\
&\label{x_pb22}x_{ij} \in [0,1]~, \quad \forall i \in \mathcal{L},~ \forall j \in \mathcal{M}\cup\mathcal{SC}
\end{align}
\end{subequations}
where all $r_{ij}$'s can be computed beforehand. This problem is a convex program which can be solved globally to the desired precision in polynomial time~\cite{nimr}.

We now show that a lower bound on the optimal value of the objective function in ${\textbf{P}}_{delay}(2,K)$ can be computed by solving ${\textbf{Q}}_{delay}$. A sketch of the proof is provided in the appendix.

\begin{thm}\label{d3_lem}
Given $\bar{\rho}$, $K$, $\overline{\alpha}$, $\lambda$, and $F$, let $q^\star$ denote the optimal value of the objective function in ${\textbf{Q}}_{delay}$. We have:
\[\left(\frac{-(|\mathcal{SC}|+|\mathcal{M}|)+q^\star}{{\sum_{i\in \mathcal{L}}{{\alpha}_i \lambda} }}\right) \le p^\star\]
where $p^\star$ denotes the optimal value of the objective function in ${\textbf{P}}_{delay}(2,K)$.
\end{thm}

Using this property, we can now work with ${\textbf{Q}}_{delay}$ to compute a lower bound on the minimum average system delay for large Hetnets. Although we are unable to verify the tightness of these bounds analytically, we will numerically verify the tightness of the computed lower bound by finding a feasible solution to the problem ${\textbf{P}}_{delay}(2,K)$, and then comparing the average delay $d_2(\{T_i\})$ for this feasible solution with the computed lower bound. We will use the simple association rules, discussed in Section \ref{numerical_dynamic}, to generate feasible solutions.

In summary, we have developed ways to compute exactly the maximum achievable rate and the maximum per class delay and a lower bound on the average system delay for different RA schemes by optimizing the UA and the parameter of the RA. Recall that, the purpose of this study is twofold: First, we want to compare, using our \emph{queueing-based approach},  the three RA schemes (i.e., \emph{CCD}, \emph{PSD}, and \emph{OD}) in terms of maximum achievable arrival rate, maximum average delay per class, and average system delay, and to study how different simple association rules perform as compared to the optimal UA solutions for the three RA schemes\footnote{\textcolor{black}{\label{UA_class}The optimal UA in the class of non-load based UAs.}}. Second, we want to compare the \emph{snapshot and queueing-based  approaches} qualitatively, and to draw conclusions on the robustness of the engineering insights obtained via the two approaches. In Section \ref{numerical_dynamic}, we compare different combinations of UA and RA schemes using the \emph{queueing-based  approach}, and in Section \ref{comp_App}, we provide a qualitative comparison between the two modeling approaches.

\section{Numerical Results For The Queueing-based  Approach} \label{numerical_dynamic}
We start by describing the simple association rules that we are going to study and compare with the optimal ones (one optimal UA per problem)\textsuperscript{\ref{UA_class}}.

\subsection{User Association Rules}\label{simple_rule_dynamic}
We have assumed that users arrive, download a file, and depart when their files have been downloaded completely. Network operators typically associate users using some simple association rules. The most common rules use physical layer parameters to determine the BS each user should associate to. We study the following user association rules \textcolor{black}{that belong to the class of UA schemes considered in this study}:
\begin{enumerate}
\item \textbf{Best SINR:} A user at location $i\in \mathcal{L}$ associates with BS $j^\star$ that provides
 the highest per channel SINR, i.e., $j^\star = \arg \max_{j \in  \mathcal{M}\cup\mathcal{SC}} {\{\gamma_{ij}\}}$.
 \item \textbf{Range Extension (RE)~\cite{RE}:} A user at location $i$ associates with BS $j^\star = \arg \min_{j \in  \mathcal{M}\cup\mathcal{SC}} {\{\delta_{ij}\}}$
where $\delta_{ij}$ is the path loss from BS $j$ to location $i$\footnote{The association rule \emph{RE} is not the same as the rule called ``Range Expansion" which adds a bias to the reference signal power received from small cells to artificially extend their coverage areas \cite{a_survey_3GPP}, \cite{RExp}.}.

\item \textbf{Small-cell First (SCF)~\cite{pimrc2011}:} A user at location $i$ associates with small cell ${j^\star = \arg \max_{j \in \mathcal{SC}} {\{\gamma_{ij}\}}}$ as long as ${\gamma_{i{j^\star}}}> \beta$ where $\beta$ is a tuning parameter. If $\max_{j \in \mathcal{SC}} {\{\gamma_{ij}\}} < \beta$, the user at location $i$ associates with the BS that gives the maximum SINR.
 \end{enumerate}

For each of these rules, we can compute the values of $x_{ij}$ for all locations $i$ and BSs $j$ \textcolor{black}{when we fix the resource allocation scheme and its parameters (i.e., either \emph{CCD} or, \emph{OD} or \emph{PSD} with a given $K$)} . Note that the RA scheme determines the power per channel, and hence impacts the interference. Given a combination of a UA and a RA, and $\overline{\alpha}$, the in-homogeneity vector, we compute the maximum achievable arrival rate, and for a given $\lambda$ the maximum average delay per class, and the average system delay. We can also find the optimal $K$ (for OD and PSD), by computing our performance metrics iteratively for all possible values of $K$, and then selecting the best solution. Next, we compare different combinations of UA and RA schemes using the queueing-based approach.

\subsection{Parameter Settings}
\textcolor{black}{We consider a system composed of 19 macro cells. Each macro cell is overlaid with four small cells. The system has an inter-cell distance of 500 m. We use a wrap around technique in which a hexagonal cell layout  with radius $R=500/\sqrt{3}~m$ is considered (see Fig. \ref{system_model162})}. The macro BSs are located at the center of the cells while the BSs of the small cells are located around the macro BS at a distance $d=230~m$ and are placed symmetrically from the center.
We assume that the system is an OFDM system with $300$ sub-channels. We consider a reuse factor of ``three", i.e., each macro BS has access to  $M=100$ sub-channels. We also take $\bar{\rho}=0.95$.

\textcolor{black}{We assume that there are $L=2000\times19$ possible user locations in the system area, and take $\alpha_i=1/(2000\times19)$ for all $i\in\mathcal{L}$ except when otherwise specified}. We assume that users arriving to the system download files whose sizes are i.i.d. random variables of mean $F=10^6$ bits.


\begin{center}
\begin{table}
    \caption{Physical Layer Parameters}
    \label{table_phy_dynamic}
    \begin{center}
        \resizebox {0.4\textwidth }{!}
        { 
        \begin{tabular}{|c|c||c|c|}
        \hline
        ${\text{Noise~Power}}$               & $-174~ \frac{\text{dBm}}{\text{Hz}}$ & $T_{\text{subframe}}$ & ${1~\text{ms}}$    \\
        \hline
        $\widehat{P}_p$  & ${30 ~\text{dBm}}$    & $\widehat{P}_m$  & ${46 ~\text{dBm}}$       \\
        \hline
        ${\text{UE Ant. Gain}}$   & ${0~ \text{dB}}$     &     ${\text{Sub-channel Bandwidth}}$          & ${180~\text{KHz}}$  \\
        \hline
       ${\text{Shadowing s.d.}}$ & ${8~\text{dB}}$       & ${\text{Penetration Loss}}$ & ${20~ \text{dB}}$ \\
        \hline
       $\text{SC}_{\text{ofdm}}$ & $12$       & $\text{SY}_{\text{ofdm}}$               & $14$ \\
        \hline
        $\text{Path Loss Small Cell}$ & \multicolumn{3}{|c|}{$140.7+36.7~{\text{log}}_{10}(d/1000),~d \ge 10m$}\\
        \hline
        $\text{Path Loss Macro}$ & \multicolumn{3}{|c|}{$128+37.6~{\text{log}}_{10}(d/1000),~d \ge 35m$}\\
        \hline
        \end{tabular}}
    \end{center}
\end{table}
\end{center}

\begin{table*}
    \caption{Modulation and Coding Schemes-LTE \cite{Christian}}
    \label{disc_rate_dynamic}
    \begin{center}
       \resizebox {0.8\textwidth }{!}
        {\begin{tabular}{|c|c|c|c|c|c|c|c|c|c|c|c|c|c|c|c|}
        \hline
        ${\text{SINR thresholds (in dB)}}$  & -6.5 & -4 & -2.6 & -1 & 1 & 3 & 6.6 & 10 & 11.4 & 11.8 & 13 & 13.8 & 15.6 & 16.8 & 17.6\\
        \hline
        ${\text{Efficiency (in bits/symbol)}}$  & 0.15 & 0.23 & 0.38 & 0.60 & 0.88 & 1.18 & 1.48 & 1.91 & 2.41 & 2.73 & 3.32 & 3.90 & 4.52 & 5.12 & 5.55 \\
        \hline
        \end{tabular}}
    \end{center}
\end{table*}

The physical layer parameters are based on the 3GPP evaluation
methodology \cite{parameters}. The parameters are given in Table \ref{table_phy_dynamic}. We use the SINR
model introduced in Section \ref{sys_dynamic}, and a channel model that accounts for
path loss and slow fading. Slow fading is modeled by a log-normal
shadowing with standard deviation of 8 dB, and path losses for small cells
and macro BSs are given in Table \ref{table_phy_dynamic}. The system is using an adaptive modulation and coding scheme with discrete rates.
The mapping between the SINR and the efficiency (in bits/symbol) for the modulation and coding schemes (MCS) in LTE is shown in
Table \ref{disc_rate_dynamic}. The bit rate obtained by a user that has a SINR between level $\ell$
and level $\ell+1$ is $r = \frac{\text{SC}_{\text{ofdm}}~
\text{SY}_{\text{ofdm}}}{T_{\text{subframe}}} e_{\ell}$
where $e_{\ell}$ is the efficiency ($\text{bits} / \text{symbol}$)
of the corresponding level $\ell$, $\text{SC}_{\text{ofdm}}$ is the
number of data subcarriers per sub-channel bandwidth,
$\text{SY}_{\text{ofdm}}$ is the number of OFDM symbols per
subframe, and $T_{\text{subframe}}$ is the subframe duration in time
units. These parameters are given in Table
\ref{table_phy_dynamic}.
The association rule ``Small-cell First" has a tuning parameter $\beta$. We assume that
$\beta$ can take any one of the SINR threshold values shown in Table \ref{disc_rate_dynamic}.

We compute the optimal values of our performance metrics for different RA schemes as well as the values of our metrics for different combinations of UA and RA schemes for 50 networks. A network corresponds to the random realization of the shadowing \textcolor{black}{coefficients for the $L=2000\times19$ locations} from all the BSs in the multi-tier system. In contrast to the \emph{snapshot approach}, we do not need to randomly drop users in the system area, and compute the average results over multiple realizations. We only need to consider multiple network realizations corresponding to different shadowing environments. In this section, we show the averaged results over the 50 network realizations.

\begin{figure}[t]
\begin{center}
\includegraphics[width=3in]{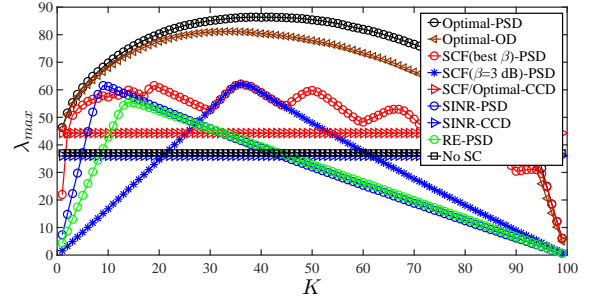} \caption{PSD, OD, and CCD: The maximum achievable arrival rate as a function of $K$ for configuration 1.}\label{stability_r1}
\end{center}
\end{figure}
\begin{figure}[t]
\begin{center}
\includegraphics[width=3in]{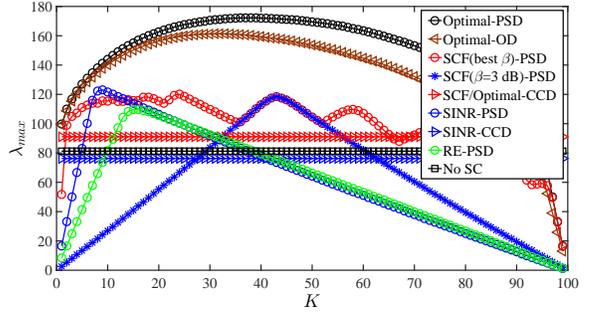} \caption{PSD, OD, and CCD: The maximum achievable arrival rate as a function of $K$ for configuration 2.}\label{stability_r1_NUD}
\end{center}
\end{figure}

\subsection{Comparison Results of OD, CCD, and PSD}
We first focus on \emph{network stability}, and compare the different resource allocation schemes in terms of the maximum achievable arrival rate ${{\lambda}}_{max}^\star(X,\overline{\alpha})$ where $X$ is the RA scheme, for two configurations. In configuration 1, \textcolor{black}{the traffic distribution is homogeneous, i.e., $\alpha_i=1/(2000\times19)$ for all $i\in\mathcal{L}$, and in configuration 2, there is a hot-spot in each cell area. Each hot-spot is a square of 150 meter in length, centered at the BS of the small cell marked by an arrow in Fig. \ref{system_model162}. We take $\alpha_i=6.58\times10^{-5}$ for user locations in the hot-spots, and $\alpha_i=1.32\times 10^{-5}$ for other locations in the system area. Note that there are 500 locations in each hot-spot.}

Figures \ref{stability_r1}-\ref{stability_r1_NUD} show  the maximum achievable arrival rate ${{\lambda}}_{max}^\star(X,\overline{\alpha})$ for different RA schemes as well as ${{\lambda}}_{max}(X,\textbf{\textit{UA}},\overline{\alpha})$ for different combinations of UA and RA schemes. We also show the maximum achievable arrival rate for the system without small cells. The curves corresponding to the optimal solution to ${\textbf{P}}_{s}$ (corresponding to the optimal user association) are labeled \emph{Optimal} in the figures. The results show that:
\begin{itemize}
\item The comparison of the highest achievable arrival rate (using the optimal solution) between the system with and without small cells (``No SC" in the figures) shows that small cells can significantly increase the maximum achievable arrival rate. We saw gains (with respect to the system without small cells) in maximum achievable arrival rate in the range of 110\% to 130\%.
\item \emph{PSD} and \emph{OD} work significantly better than \emph{CCD} for almost all values of $K$. \emph{PSD} performs better than \emph{OD}.
\item For \emph{PSD} and \emph{OD}, the association rules \emph{Small-cell First}, \emph{Range Extension}, and \emph{Best SINR} perform almost the same with a slight advantage for \emph{Small-cell First}. The performance of these rules is far from optimal. This can be explained by the fact that \textcolor{black}{even if the optimal UA does not perform dynamic load balancing, it does perform a static one, i.e., it finds for each value of $K$ the best UA mapping while the practical UAs impose a mapping for each value of $K$.} Moreover, if the value of the parameter $K$ is not chosen carefully, all these simple association rules can do worse than the system without small cells.
\item The presence of a hot-spot does not seem to impact the trends discussed above.
\end{itemize}

\begin{figure}
\begin{center}
\includegraphics[width=3in]{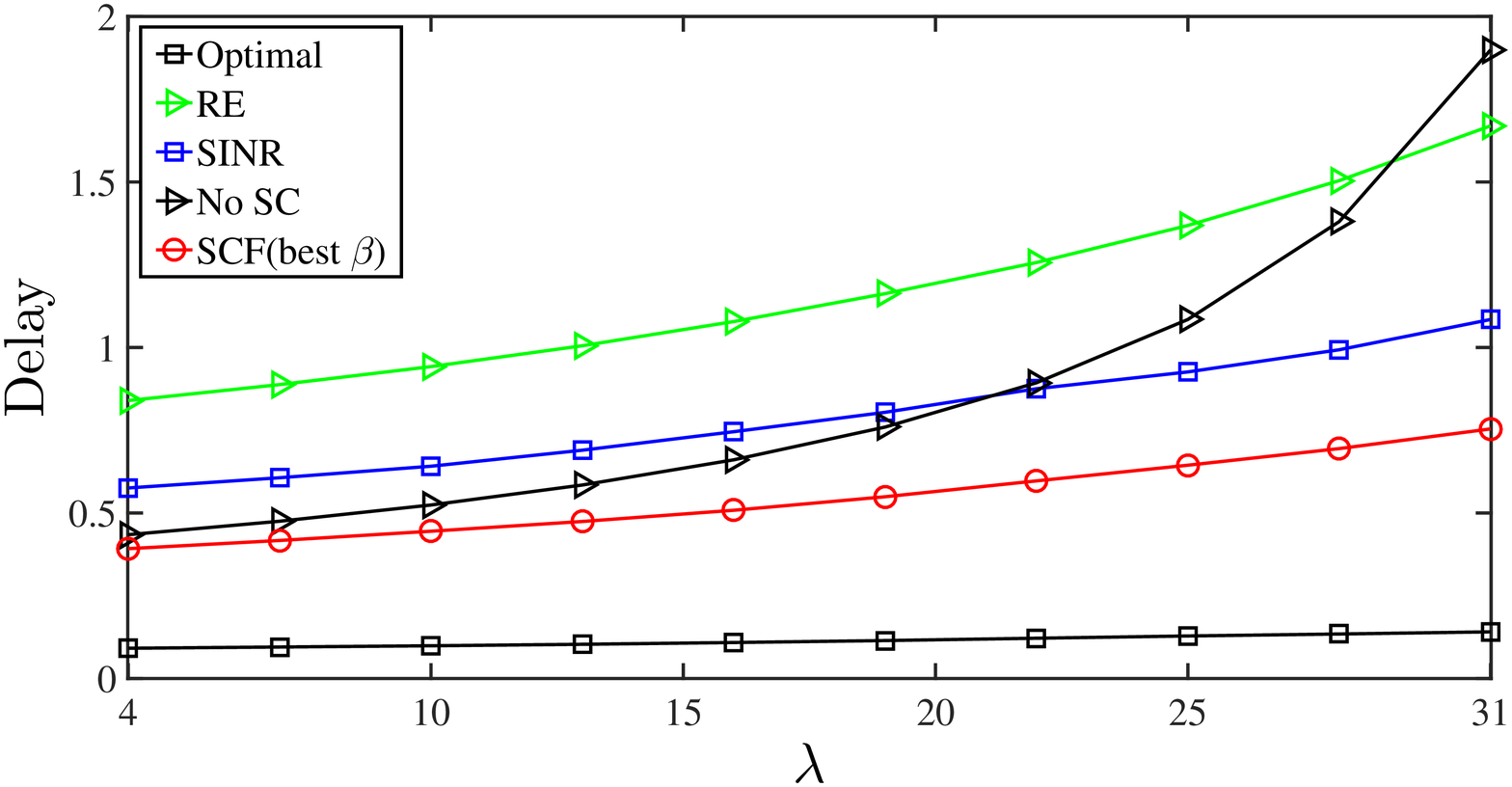}\caption{PSD: The maximum average delay per class as a function of $\lambda$ when $F=10^6$ bits. We choose the best values of $K$ and $\beta$ for each value of $\lambda$.}\label{delay_max_l}
\end{center}
\begin{center}
\includegraphics[width=3in]{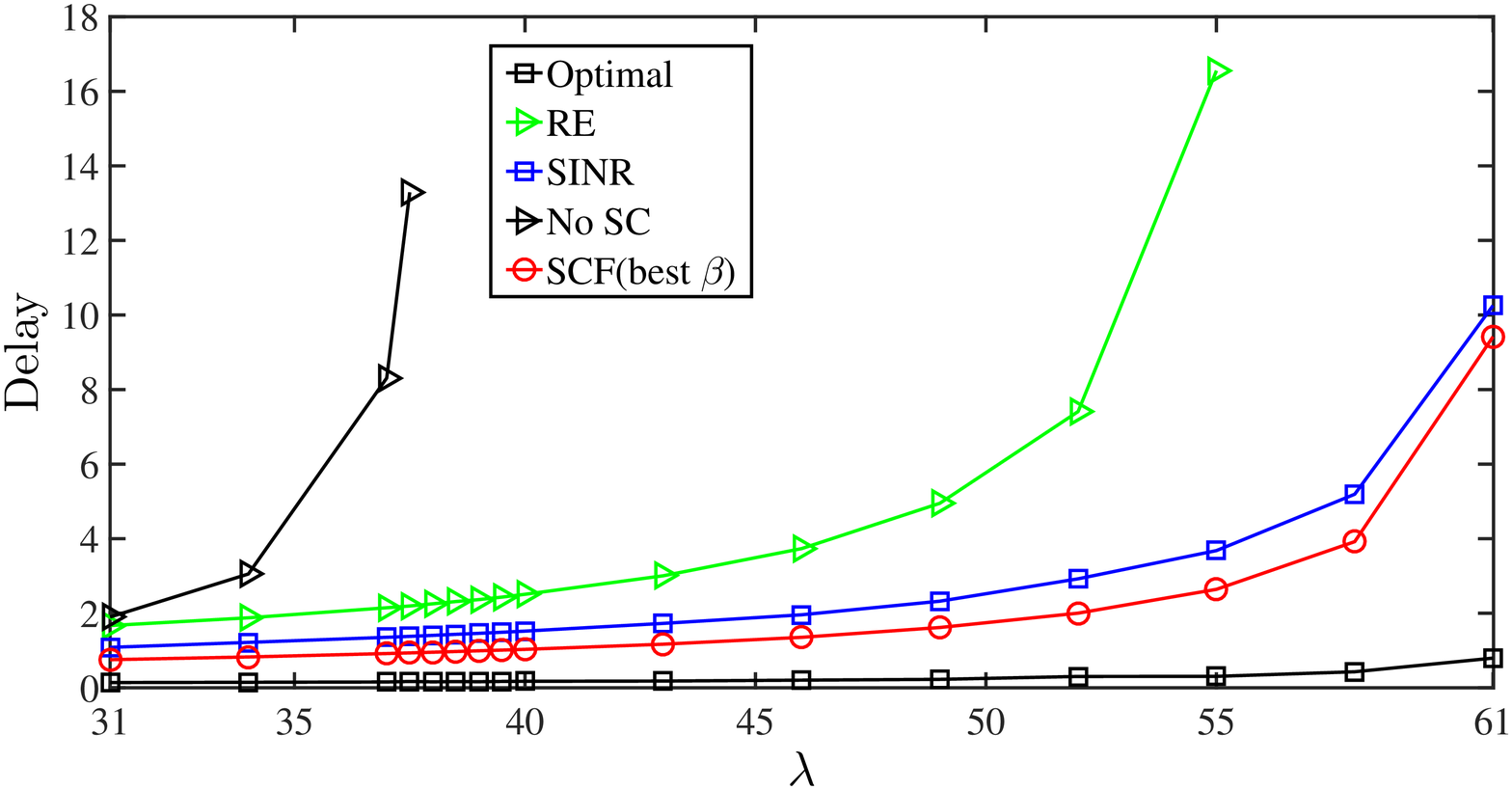}\caption{PSD: The maximum average delay per class as a function of $\lambda$ when $F=10^6$ bits. We choose the best values of $K$ and $\beta$ for each value of $\lambda$.}\label{delay_max_h}
\end{center}
\end{figure}

\subsection{In Depth Study of PSD}
We now study \emph{partially shared deployment}, the best of the three resource allocation schemes, in more details. To do so, we first focus on the maximum average delay per class, and then study the average system delay.

\subsubsection{Maximum average delay per class}
We select the maximum average delay per class as our delay metric, and compare the delay performance of the simple UA rules with the optimal delay performance as a function of $\lambda$, the total arrival rate into the cell area. We fix the arrival rate $\lambda$, and compute the optimal solution to the problem ${\textbf{P}}_{delay}(1)$ with the precision $\epsilon=0.02$, and the corresponding maximum average delay per class for each UA rule. For each UA rule, we select the value of $K$ which results in the lowest maximum average delay per class. The results for two non-overlapping ranges of $\lambda$ are shown in Figures \ref{delay_max_l}-\ref{delay_max_h}. The curve corresponding to the optimal solution is labeled \emph{Optimal} in the figures. The results show that:
\begin{itemize}
\item The association rules \emph{Small-cell First} and \emph{Best SINR} perform almost the same with a slight advantage for \emph{Small-cell First}, and they work significantly better than \emph{Range Extension} for all values of $\lambda$ when we select $\beta$ and $K$ carefully.
\item The association rule \emph{Small-cell First} performs better than the system without small cels for all values of $\lambda$. However, the rules \emph{Best SINR} and \emph{Range Extension} do not always perform better than the system without small cells (especially for low values of $\lambda$).
\item None of the simple rules are performing very well for high values of $\lambda$.
\item The association rule \emph{Range Extension} is not performing better than the system without small cells for low values of $\lambda$.
\end{itemize}

\begin{figure}
\begin{center}
\includegraphics[width=3in]{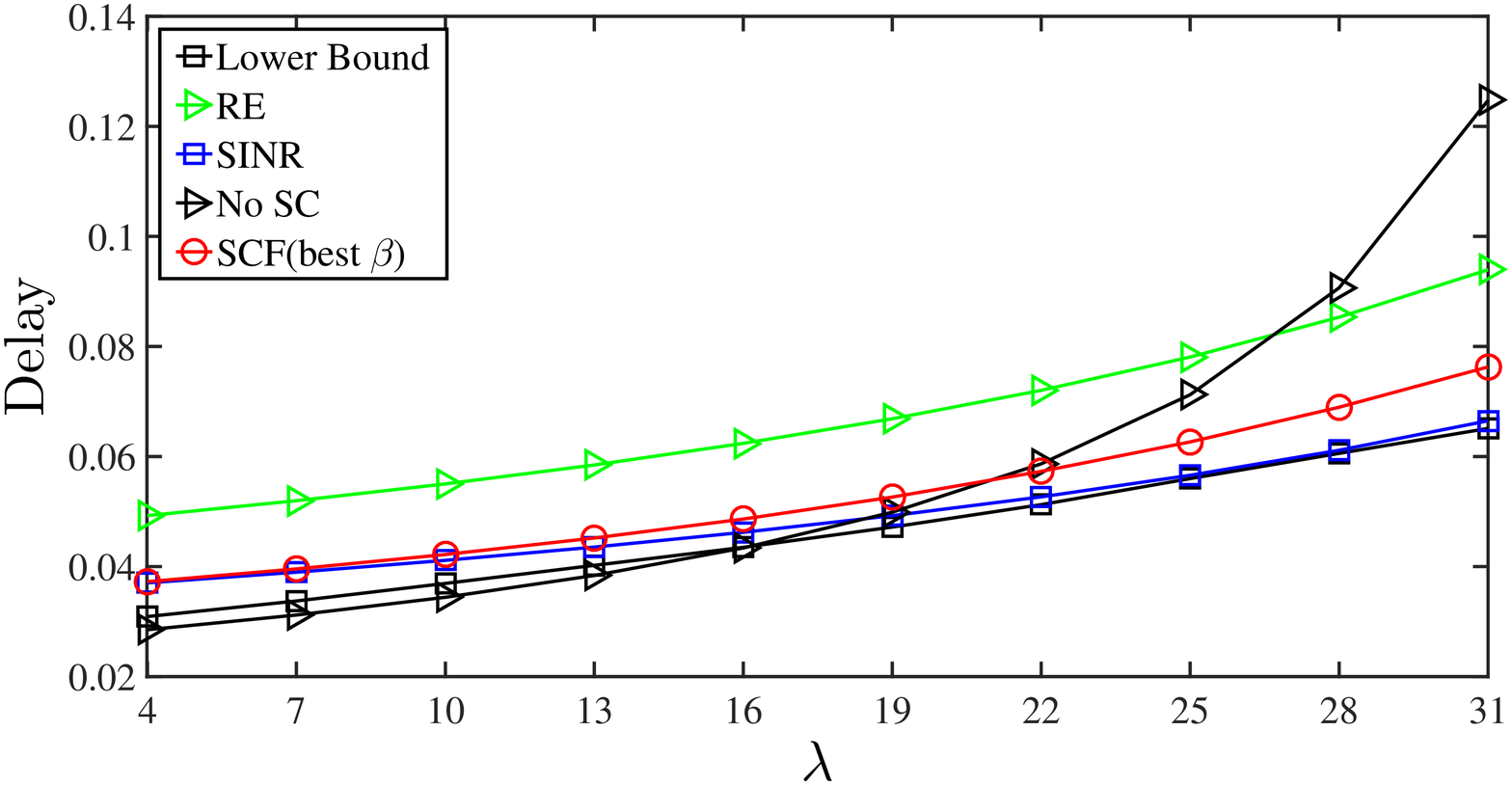}\caption{PSD: The average system delay as a function of $\lambda$ when $F=10^6$ bits. We choose the best values of $K$ and $\beta$ for each value of $\lambda$.}\label{delay_ave_l}
\end{center}
\begin{center}
\includegraphics[width=3in]{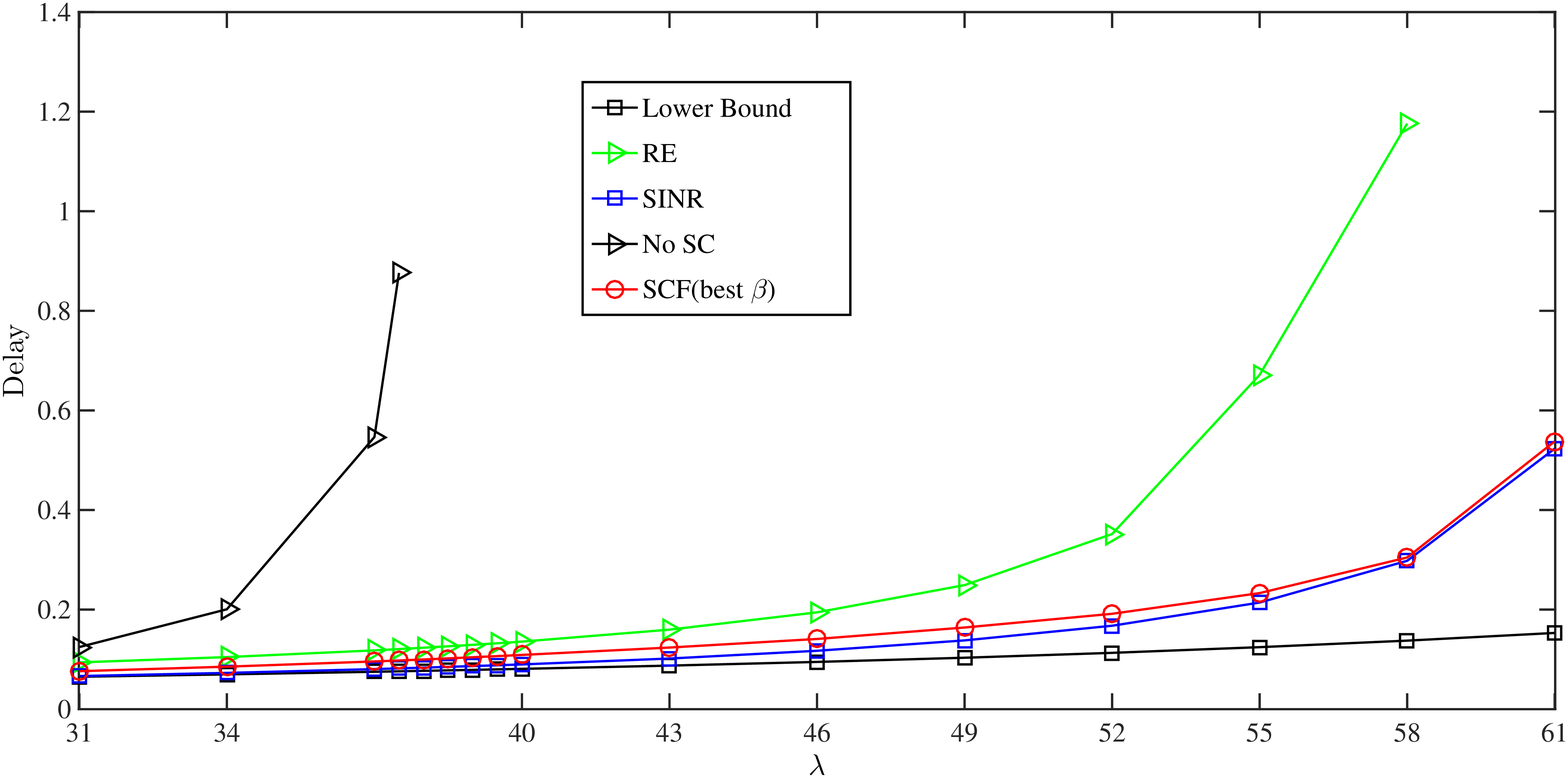}\caption{PSD: The average system delay as a function of $\lambda$ when $F=10^6$ bits. We choose the best values of $K$ and $\beta$ for each value of $\lambda$.}\label{delay_ave_h}
\end{center}
\end{figure}

\subsubsection{Average system delay}
We now compare the average system delay of the simple UA rules with the lower bound on the optimal average system delay as a function of $\lambda$. We fix the arrival rate $\lambda$, and compute the lower bound of the optimal joint user association and resource allocation problem ${\textbf{P}}_{delay}(2)$ for \emph{PSD}, and the average system delay for each UA rule when the RA parameter $K$ is computed optimally. To check the tightness of the computed lower bound, we compare it with the average system delay of the simple UA rules. The results for two non-overlapping ranges of $\lambda$ are shown in Figures \ref{delay_ave_l}-\ref{delay_ave_h}. The curve corresponding to the lower bound is labeled \emph{Lower Bound} in the figures. The results show that:
\begin{itemize}
\item The UA rule \emph{Best SINR} is performing very well since the average delay of \emph{Best SINR} is very close to the computed lower bound for a large range of $\lambda$ when the RA parameter $K$ is chosen optimally. This validates our relaxation approach since an integer solution to the proposed problem yields almost the same average delay as the solution of the relaxed problem. This observation also shows that \emph{Best SINR} is a good UA rule for minimizing the average system delay if we choose $K$ optimally.
\item The association rules \emph{Small-cell First} and \emph{Best SINR} perform almost the same with a slight advantage for \emph{Best SINR}, and they work significantly better than \emph{Range Extension} for all values of $\lambda$ when we select $\beta$ and $K$ carefully.
\item The comparison of the delay performance (using \emph{the lower bound}) between the system with and without small cells shows that small cells can increase the average system delay when the arrival rate $\lambda$ is relatively low (less than 19 users per second). This shows the critical impact of the interference caused by the small cells on the delay performance for small values of $\lambda$.
\end{itemize}

\textcolor{black}{{\bf In summary}, the user association that works well for the three metrics is \emph{Small-cell First}. Note also that if the user arrival rate  is very low, the average delay is better when there is no small cells if we believe our model, but this is questionable in view of our assumption A.7. As mentioned earlier,  dynamic inter-cell interference and load-coupling can play a significant role in the system when the user arrival rate is low. In such a case, detailed simulations are needed to validate our results.}


\section{Comparison Between The two Modeling Approaches}\label{comp_App}
The \emph{snapshot} and \emph{queueing-based} approaches model cellular systems under different sets of assumptions, and address different problems while they have the same overall goal to compare different combinations of UA, RA, and scheduling schemes. The \emph{snapshot} approach allows us to formulate many network utility maximization problems, and to evaluate the throughput performance of
many combinations of scheduling, power control, UA, and RA schemes over a large number of independent snapshots of
the system, but it does not allow us to capture the system dynamics. While the \emph{queueing-based} approach enables us to introduce some dynamic elements into the system model to take into account the users' arrival and departure processes, it does not allow us to evaluate the performance of different combinations of power control and scheduling schemes since processor sharing is only valid under our set of assumptions. \textcolor{black}{Moreover, we cannot decompose the set of PS queues into independent queues if we use a user association rule that takes a decision based on the load of each BS or assume that there is coordination among different BSs.}


\textcolor{black}{ The \emph{snapshot} and \emph{queueing-based }  modeling approaches are used in offline design to separate quickly the RA schemes and UA policies which are promising, from the ones which are not. These approaches to be tractable have to have some limiting assumptions. Therefore, these analytical modeling approaches should be seen as a first line of study. Doing this first level of selection with simulation is difficult due to the many options and parameters.  When the engineering insights obtained via the \emph{snapshot approach} are consistent with the insights drawn out of the \emph{queueing-based} study, we can feel confident that the insights are valid while if there are different, there is a clear need for further studies via simulation for example.}

We have compared the \emph{snapshot} and \emph{queueing-based} approaches in terms of the trends they highlight (e.g., scheme ``a" is better than scheme ``b")
to draw conclusions on the robustness of the engineering insights obtained via the two approaches.


{\bf Comparison of  the three RA schemes:}  The \emph{snapshot approach} (using our results in  \cite{df}) shows that \emph{PSD} and \emph{OD} perform significantly better than \emph{CCD} when the user association is optimal and $K$ is chosen well, irrespective of the number of users in the system. The results also show that \emph{PSD} performs better than \emph{OD}. These engineering insights are consistent with those obtained via the \emph{queueing-based approach} irrespective of the metric being used.

{\bf Small cells versus no small cells:} The \emph{snapshot approach} (using our results in  \cite{df}) shows that the system without small cells  performs worse than the system with small cells with the optimal user association except for some extreme values of $K$. The number of users in the system does not change this conclusion. The \emph{queueing-based  approach} when the metric is either the maximum achievable rate or the maximum delay per class gives similar results. However, when the metric is the average delay, the system without small cells performs better if the traffic is low.

{\bf Comparison of the association rules on \emph{PSD}:}  The \emph{snapshot approach} (using our results in  \cite{df})  shows that \emph{SCF}  performs better than \emph{Best SINR} and \emph{RE} irrespective of the number of users in the cell, and that the association rules \emph{Best SINR} and \emph{RE} perform almost the same for all possible values of the number of users. It also shows that \emph{SCF} is quasi-optimal when the number of users in the system is large. Our numerical results also show that the system with small cells operating with the simple association rules performs better than the system without small cells for a large range of $K$.

The results obtained via the \emph{queueing-based approach} show that none of the association rules perform extremely well and that the association rules \emph{SCF} and \emph{Best SINR} perform almost the same in terms of the maximum achievable arrival rate, and that they perform better than \emph{RE} when $K$ is optimized. These results show that the engineering insights on the association rules for \emph{PSD} obtained via the \emph{snapshot approach} are not always consistent with the insights drawn out of the \emph{queueing-based} study. \textcolor{black}{The fact that the two approaches give us different trends should be taken as a strong indication that a further study, maybe via simulations, is needed.}

\section{Conclusions} \label{concl}
\textcolor{black}{We have proposed a tractable queueing-based  framework to analyze and compare different combinations of UA and RA schemes in an offline (and centralized) fashion}.  We have chosen three different performance metrics: the maximum achievable arrival rate, the average system delay, and the maximum average delay per class, and formulated three different UA problems to optimize our performance metrics under spatially homogeneous and in-homogeneous traffic distributions.

In this study, we have compared the two modeling approaches to draw conclusions on the ``robustness" of the engineering insights obtained via the \emph{snapshot} and \emph{queueing-based } approaches. Our numerical results indicate that the engineering insights on the RA schemes obtained via the \emph{snapshot approach} are valid in a dynamic context, and vice versa. However, the comparative study of the association rules in \emph{Partially shared deployment} shows the lack of robustness of certain insights drawn out of the \emph{snapshot approach}.

The engineering insights obtained via the \emph{snapshot approach} indicate that \emph{Small-cell First} performs better than the existing rules, and that it is quasi-optimal. However, the numerical results obtained out of the \emph{queueing-based } study indicate that \emph{Small-cell First} performs significantly better than the other rules only for edge users (it performs as well as the other rules for other users), and that the conventional association rule (i.e., \emph{Best SINR}) performs relatively well except for edge users. Our numerical results, obtained from the \emph{queueing-based  approach}, also indicate that UA rules that do not take load balancing into account, do not perform very well in practical systems.

\section{Appendices}
\section{Proof of Theorem 1}
To prove the theorem, we need to show that an optimal solution to ${\textbf{P}}_{s}'(K)$ is optimal to ${\textbf{P}}_{s}(K)$, and vice versa.

Let $(\{x_{ij}^\star\},{{\lambda}}^\star)$ denote the optimal solution to ${\textbf{P}}_{s}(K)$. We can easily verify that ${{\lambda}}^\star \left(\max_{j \in \mathcal{M}\cup\mathcal{SC}}{\{\sum_{i\in\mathcal{L}}{x_{ij}^\star \frac{\alpha_i F}{K_j r_{ij}}}\}}\right)=\bar{\rho}$; otherwise, we will get a contradiction with the assumption that ${{\lambda}}^\star$ is the optimal value of ${{\lambda}}$ in ${\textbf{P}}_{s}(K)$. Now, let us assume that $\{x_{ij}^\star\}$ is not an optimal solution to ${\textbf{P}}_{s}'(K)$. Therefore, there exists $\{y_{ij}^\star\}$ that is optimal for ${\textbf{P}}_{s}'(K)$, and $\{y_{ij}^\star\}$ satisfies the following inequality:
\[\max_{j \in \mathcal{M}\cup\mathcal{SC}}{\left\{\sum_{i\in\mathcal{L}}{y_{ij}^\star \frac{\alpha_i F}{K_j r_{ij}}}\right\}} <
\max_{j \in \mathcal{M}\cup\mathcal{SC}}{\left\{\sum_{i\in\mathcal{L}}{x_{ij}^\star \frac{\alpha_i F}{K_j r_{ij}}}\right\}}~.\]

Let us choose $\lambda^\prime$ as follows:
\[\lambda^\prime=\frac{\bar{\rho}}{\max_{j \in \mathcal{M}\cup\mathcal{SC}}{\left\{\sum_{i\in\mathcal{L}}{y_{ij}^\star \frac{\alpha_i F}{K_j r_{ij}}}\right\}}}~.\]
We can easily verify that $(\{y_{ij}^\star\},\lambda^\prime)$ is a feasible solution for ${\textbf{P}}_{s}(K)$, and that $\lambda^\prime>\lambda^\star$. This contradicts the assumption that $\lambda^\star$ is optimal for ${\textbf{P}}_{s}(K)$. Therefore, $\{x_{ij}^\star\}$ is an optimal solution to ${\textbf{P}}_{s}'(K)$. By following the same argument, we can show that an optimal solution to ${\textbf{P}}_{s}'(K)$ is optimal to ${\textbf{P}}_{s}(K)$. This completes the proof. \qed

\section{Proof of Theorem 2}
Given $\overline{\alpha}$, $\lambda$, $K$, and $F$, ${\textbf{P}}_{delay}(1,K)$ is defined as follows:
\begin{subequations}
\begin{align}
{\textbf{P}}_{delay}(1,K):~~&\min_{\{x_{ij}\},\{\rho_j\},\{T_i\},L}  \quad {L}\nonumber\\
\text{subject to}  & \quad (\ref{stability_cond_pb}), (\ref{UA_const_pb}), (\ref{delay_per_loc_pb})-(\ref{x_pb})\nonumber\\
\label{K_2}& T_i \le L,~~\forall i \in \mathcal{L}
\end{align}
\end{subequations}
where all $r_{ij}$'s can be computed beforehand.

The structure of ${\textbf{P}}_{delay}(1,K)$ is
such that we can reformulate it as follows. Note that all $x_{ij}$'s are binary variables, and that
$\sum_{j \in \mathcal{SC}}{x_{ij}}=1$ for all user locations $i$. Therefore, for each user location $i\in\mathcal{L}$, there exists only one
value of $j$, call it $\widehat{j}$, for which $x_{i\widehat{j}} = 1$ (i.e., $x_{ij} = 0$, $\forall j \neq \widehat{j}$). Therefore, we can easily show that the constraints (\ref{delay_per_loc_pb}) and (\ref{K_2}) in ${\textbf{P}}_{delay}(1,K)$ are equivalent to the following constraint:
\[
{x_{ij} \frac{F}{(1-\rho_j)K_j r_{ij}}} \le L ,~~\forall i \in\mathcal{L},~ \forall j\in \mathcal{M}\cup\mathcal{SC}
\]
Using this property, ${\textbf{P}}_{delay}(1,K)$ can be
reformulated as follows:
\begin{subequations}
\begin{align}
{\textbf{P}}^\prime_{delay}(1,K):~~&\min_{\{x_{ij}\},\{\rho_j\},L}  \quad {L}\nonumber\\
\text{subject to}&\quad (\ref{stability_cond_pb}), (\ref{UA_const_pb}), (\ref{rho_pb})-(\ref{x_pb})\nonumber\\
x_{ij} \frac{F}{(1-\rho_j)K_j r_{ij}} &\le L ,~~\forall i \in\mathcal{L},~ \forall j\in \mathcal{M}\cup\mathcal{SC} \nonumber
\end{align}
\end{subequations}

Let $p^\star$ denote the optimal value of the objective function in ${\textbf{P}}_{delay}(1,K)$. Now, we can easily verify that if ${\textbf{P}}^{\prime}_{delay}(t)$ is feasible for a given value $t>0$, then we have $p^\star \le t$; otherwise, we have $p^\star>t$. This completes the proof. \qed

\section{Proof of Theorem 3}
Given $\overline{\alpha}$, $\lambda$, $K$, and $F$, ${\textbf{P}}_{delay}(2,K)$ is defined as follows:
\begin{subequations}
\begin{align}
{\textbf{P}}^\prime_{delay}(2,K):~~&\min_{\{x_{ij}\},\{\rho_j\},\{T_i\}}  \quad {\frac{\left({\sum_{i\in \mathcal{L}}{{\lambda}_i T_i}}\right)}{\left({\sum_{i\in \mathcal{L}}{{\lambda}_i} }\right)}}\nonumber\\
\text{subject to}& \quad (\ref{stability_cond_pb}), (\ref{UA_const_pb}), (\ref{delay_per_loc_pb})-(\ref{x_pb})\nonumber
\end{align}
\end{subequations}
where all $r_{ij}$'s can be computed beforehand.

Using eq.~\ref{delay_per_loc_def}, we can verify that the average system delay $d_2(\{T_i\})$ is equal to
\begin{subequations}
\begin{align}
&\frac{\left({\sum_{i\in \mathcal{L}}{{\lambda}_i T_i}}\right)}{\left({\sum_{i\in \mathcal{L}}{{\lambda}_i} }\right)}=\frac{1}{\left({\sum_{i\in \mathcal{L}}{{\lambda}_i} }\right)}\times \sum_{j \in \mathcal{M}\cup\mathcal{SC}}{\frac{\rho_j}{1-\rho_j}}\nonumber\\
&=\frac{1}{\left({\sum_{i\in \mathcal{L}}{{\lambda}_i} }\right)}\times\left(-(|\mathcal{SC}|+|\mathcal{M}|)+\sum_{j \in \mathcal{M}\cup\mathcal{SC}}{\frac{1}{1-\rho_j}}\right)~.\nonumber
\end{align}
\end{subequations}
Therefore, minimizing the average system delay is equivalent to minimizing $\sum_{j \in \mathcal{M}\cup\mathcal{SC}}{\frac{1}{1-\rho_j}}$ since ${\lambda}_i$s are given beforehand.

The optimal value, $p^\star$, of the objective function in ${\textbf{P}}_{delay}(2,K)$ is equal to
$({-(|\mathcal{SC}|+|\mathcal{M}|)+\widehat{q^\star}})/({{\sum_{i\in \mathcal{L}}{{\lambda}_i} }})$
where $\widehat{q^\star}$ denotes the optimal value of the objective function in ${\textbf{Q}}_{delay}$. By relaxing the integrality constraint on the $(x_{ij})$'s, we can compute a lower bound on the optimal value of the objective function in ${\textbf{P}}_{delay}(2,K)$. This completes the proof. \qed

\end{document}